\documentstyle[11pt]{article}
\newcommand{\g}{{\rm g}}
\def\quarter{\mbox{\small{$\frac{1}{4}$}}}
\def\beq{\begin{equation}}
\def\eeq{\end{equation}}
\def\bea{\begin{eqnarray}}
\def\eea{\end{eqnarray}}
\def\eqref#1{eq.(\ref{eq:#1})}
\def\eref#1{(\ref{eq:#1})}
\def\eqlab#1{\label{eq:#1}}
\newcommand{\FMslash}[1]{#1 \hspace{-1.9mm} \slash}

\def\half{\mbox{\small{$\frac{1}{2}$}}}
\def\thalf{\mbox{\small{$\frac{3}{2}\,\,$}}}
\def\lag{{\cal L}}

\def\cO{{\cal O}}
\def\N{{\rm N}}

\def\al{\alpha}
\def\be{\beta}
\def\ga{\gamma} 
 \def\De{{\it\Delta}}
\def\Psii{{\it\Psi}}

\def\la{\lambda}  

  \def\Th{{\it\Theta}}

\def\pa{\partial}

\def\nonu{\nonumber}
\def\pa{\partial}
\def\ra{\rightarrow} 

\begin{document}
\begin{titlepage}
\begin{center}
             \vspace{-4cm}
        \hfill {\normalsize {\bf THU--94/21}}\\
        \hfill {\normalsize {\bf hep-ph 9421321}}\\[3cm]
{\LARGE \bf On the interaction of spin $\frac{3}{2}$  particles}\\[1.2cm]
{\Large Vladimir Pascalutsa}\\[0.7cm]
{\it Kernfysisch Versneller Instituut, University of Groningen, Zernikelaan
25, \\
9747 AA Groningen, The Netherlands\\[0.5cm]
Institute for Theoretical Physics, University of Utrecht,
Princetonplein 5,\\
 P.O. 80006, 3508 TA Utrecht, The Netherlands\footnote{Present address.}}
\date{December 7, 1994}
\end{center}

\bigskip
\bigskip

\begin{abstract} Lagrangian of a massive particle with spin \thalf is
considered in the Rarita-Schwinger formalism. We discuss implications
of the contact- and the gauge-transformation on the physical
content of free and interacting theories. It is shown that
the ``contact-invariance'' is trivial and has no physical relevance.
The requirement of the gauge-invariance of the free lagrangian
fixes the off-shell lower-spin content of the spin-\thalf field.
The gauge-invariance of the interaction lagrangian insures that 
these lower-spin components are not coupled, and (probably 
always) leads to theories consistent with local causality. 
  \end{abstract}
\bigskip
\end{titlepage}

Formulating a free massive spin-\thalf theory in the vector-spinor
representation of Rarita and Schwinger \cite{Rar41}, one usually finds the following form for the lagrangian (see
e.g.\ ref.\cite{Muk89}), \begin{eqnarray}
 \eqlab{free} {\cal L}_{\rm free}&=&\bar{\psi}_{\al}\,
\Lambda^A_{\alpha \beta} \,\psi_{\be}, \nonu \\ 
\Lambda^A_{\alpha
\beta}&=&(i \FMslash{\partial} - M )\, {\rm g}_{\alpha
\beta}+iA\,(\gamma_{\alpha}\partial_{\beta}+
\gamma_{\beta}\partial_{\alpha}) \\
&& + \mbox{$\frac{i}{2}$}(3A^{2}+2A+1)\,
\gamma_{\alpha}\FMslash{\partial }\gamma_{\beta}
+(3A^2+3A+1)M\,\gamma_{\alpha}\gamma_{\beta}, \nonu
\end{eqnarray}
 where  $\psi_{\al}$ is a sixteen component Lorentz vector-spinor
field (spinor index is supressed), $M$ is the mass, and $A$ is a
parameter ($A\neq - \half,$ in order for the propagator to exist).

 Equation \eref{free} defines essentially a family of
lagrangians corresponding to different values of $A$, nevertheless, all
of them lead to the same equation of motion
\beq 
(i \FMslash{\pa} - M)\, \psi_{\al}=0,
\eeq
provided the following constrains are imposed,
\bea
\eqlab{on-shell} \partial_{\alpha}\, \psi_{\al}&=&0, \nonumber \\
\gamma_{\alpha}\, \psi_{\al}&=&0. 
\eea 
The latter constrains can be imposed only on the mass shell, and their
role is to eliminate the virtual spin-\half\ components of $\psi_{\al}$
(thus, on the mass shell,  the number of
independent components of $\psi_{\al}$ is reduced to eight,
as is appropriate for a massive, free spin-\thalf field).
\bigskip

Lagrangian \eref{free} is invariant under the so-called {\it contact}
(or {\it point}) {\it transformation}: 
\begin{eqnarray} \eqlab{ptf}
\psi_{\alpha}&\rightarrow& \psi_{\alpha}' = (\g_{\alpha \beta} + \half
b \gamma_{\alpha}\gamma_{\beta}) \psi_{\be} \equiv {\cal O}^b_{\alpha
\beta} \, \psi_{\be} \nonumber \\ A&\rightarrow&A'=\frac{A-b}{1+2b}\,
, \end{eqnarray} where $b$ is restricted only by the
condition that ${\cal O}^b$ has an inverse, that is $b \neq
-\half.$ The invariance under this transformation implies that the
physical properties of the free field are invariant with respect to
rotations in the spin $\half$ space, and (equivalently) are 
independent of the choice of the parameter $A$. 

To construct a contact-invariant
interaction one needs to assume that field $\psi_{\alpha}$ enters
the lagrangian only in combination with some tensor $\Th_{\alpha \beta}$
\cite{Pec68}. For example, $  {\cal L}_{I}=
j_\mu(x) \, \Th_{\mu \alpha}\,  \psi_{\al}(x),$
 where $j_\mu (x)$ is a source. 
Original Peccei's choice \cite{Pec68}: $\gamma_{\al}\,
\Th_{\alpha \beta} = 0$, is not the most general one, and moreover
for this case $\Th$ does not have an inverse which may lead to
bad consequences.
The most general form of $\Th$ which preserves 
the invariance is found by  Nath {\it et.\ al.\ }\cite{Nat71}:
\beq
\eqlab{separation}
\Th_{\alpha \beta}= {\rm g}_{\alpha \beta} + [ z + \half (1+4 z)A
] \gamma_{\alpha} \gamma_{\beta},
\eeq
where $z$ is an arbitrary parameter, the so-called
off-shell parameter. Note that this result can also be presented as,  $\Th_{\alpha \beta}=
(\g_{\alpha \mu} + z \gamma_{\alpha}
\gamma_{\mu})\,(\g_{\mu \beta} + \half A \gamma_{\mu} \gamma_{\beta})$.

Since
the contact invariance of the full lagrangian is obeyed
the physical observables are independent of $A$. This statement is 
the subject of the Kamefuchi-O'Raifeartaigh-Salam theorem \cite{Kam61}. 

We would like to state here that the whole discussion of the 
contact invariance is physically irrelevant and can generally be avoided.
To see this we rewrite the free lagrangian \eref{free} 
in the following way, \begin{equation}
 \eqlab{new-free} {\cal L}_{\rm free}=\bar{\psi}_{\al}\, {\cal
O}^A_{\alpha \mu } \, \Lambda_{\mu \nu} \, {\cal O}^A_{\nu \beta}
\,\psi_{\be}, \end{equation}
 where
\beq \Lambda_{\mu \nu}=(i \FMslash{\partial} -
M )\, {\rm g}_{\mu \nu} - \quarter \gamma_{\mu}\gamma_{\lambda}\,
 (i \FMslash{\partial} - M ) \, \gamma_{\lambda}\gamma_{\nu}, \eeq and
 all the $A$-dependence is  contained thus in  ${\cal
O}^A_{\alpha \beta}=\g_{\alpha \beta} + \half A \gamma_{\alpha}
\gamma_{\beta}.$
Now we can just redefine the field:
\beq \Psii_{\alpha} = {\cal O}^A_{\alpha \beta}
\,\psi_{\be}.  \eeq  
Field $\Psii_{\alpha}$ is explicitly
contact invariant, and so will be a lagrangian involving it.
Moreover, from \eqref{ptf} we can see that,
\beq
\cO^{A'}_{\al\be}= \cO^{A}_{\al\mu}\,(\cO^{b}_{\mu\be})^{-1},
\eeq
therefore the transformation on field $\Psii_{\alpha}$ amounts to
an insertion of unity operator $(\cO^{b})^{-1}\,\cO^{b}$:
\beq
\Psii_{\alpha} = {\cal O}^A_{\alpha \la}\,\psi_\la\,\ra\,
 \Psii_{\alpha}'=  \cO^{A}_{\al\mu}\,(\cO^{b}_{\mu\be})^{-1}\,\cO^{b}_{\be\la}\,
\psi_\la
\eeq 
which clearly demonstrates the triviality of the contact transformation.
\bigskip
  
Let us consider now a more (physically) interesting symmetry --- gauge
symmetry. A requirement of the invariance of the kinetic term on the
Rarita-Schwinger {\it gauge transformation} \cite{Rar41}:
\beq
\eqlab{RSGT}
\psi_{\mu}\rightarrow \psi_{\mu} + \partial_{\mu} \epsilon,
\eeq
(where $\epsilon$ is a spinor field)
fixes the form of the free lagrangian to\footnote{
This form corresponds to $A=-1$  in \eqref{free} or \eqref{new-free}.
}:
\beq
 \eqlab{freeRS}
 {\cal L}_{\rm free} =\bar{\psi}_{\mu}\, \left\{  \sigma_{\mu \nu},
(i \FMslash{\partial} - M )\right\}\, \psi_{\nu},
\eeq 
where $\sigma_{\mu \nu}=\half [\ga_\mu, \ga_\nu],$ curly brackets
denote anticommutator.\footnote{
The invariance of the kinetic term is readily checked by the identity:
$
\{  \sigma_{\mu \nu},\gamma_{\alpha} \} = 
\{  \sigma_{\nu \alpha},\gamma_{\mu} \} = \varepsilon_{\mu \nu \alpha \lambda}
\,\gamma_{\lambda} \gamma_5.
$
The mass term $M \bar{\psi}^{\mu}\,  \sigma_{\mu \nu} \,
\psi^{\nu}$ obviously destroys the invariance. 
}
 
The amount of the spin-\half\ off-shell components of field $\psi_\al$
is now fixed, but they still are allowed to couple in essentially
arbitrary way, with strength proportional
to the off-shell parameter $z$. This freedom is there and may
certainly be explored phenomenologically.  However one should bear
in mind that such an arbitrary interaction can be inconsistent with
the principle of local causality. 

We would like point out that, if the interaction lagrangian
is gauge-invariant, the troublesome spin-\half\ components do
not couple. Probably because of this reason, gauge-invariant interactions 
in many (if not all) cases lead to consistent theories.
For example, requiring gauge-invariance of the $\pi\N\De$ lagrangian,
\beq
\lag_{\pi\N\De} = g_{\pi\N\De}\; \bar{\psi}_\N\,(\pa_\al\phi_\pi)
\,\Th_{\al\be}\,\psi_{\De,\,\be} + {\rm H. c.},
\eeq
we find $z=\half$, the value which Nath {\it et.\ al.\ }\cite{Nat71}
found from the requirement of consistency with causality.
Also, the successful treatment of spin-\thalf fields in supergravity is 
due to a simple fact that both spin-\thalf and spin-2 gauge-symmetry is
realized there as invariance under the supersymmetry
transformation of graviton and gravitino.  

Generally speaking, it can be that the only consistent field theories
of higher spins are the gauge-invariant theories (where only mass terms
are allowed to break gauge invariance). A rigorous proof of
a theorem like this is certainly indispensable.
\begin{center}
\[ \star\; \star\; \star \]
\end{center}

I am grateful to  O.\ Scholten and J.A.\ Tjon 
for interesting discussions on the treatment of high-spin particles.
I would like to thank the personnel of the KVI
for kind hospitality and support during my stay in Groningen.

The work was supported by de Stichting voor Fundamenteel Onderzoek der Materie
(FOM) and a fellowship from the Netherlands Organization for International
Cooperation in Higher Education (Nuffic).


\small
\begin{thebibliography}{9}
\bibitem{Rar41} W. Rarita and J. Schwinger, Phys. Rev.{ 60} (1941) 61.
\bibitem{Muk89}  M. Benmerrouche, R.M. Davidson and N.C. Mukhopadhyay, Phys.
Rev. C{39} (1989) 2339, and references therein.
\bibitem{Pec68}  R.D. Peccei, Phys. Rev. C{176} (1968) 1812; C{181} (1969)
1902.
\bibitem{Nat71}  L.M. Nath, B. Etemadi and J.D. Kimel, Phys. Rev. D{3}
(1971) 2153.
\bibitem{Kam61}  S. Kamefuchi, L. O'Raifeartaigh and A. Salam,
 Nucl. Phys.{ 28} (1961) 529.

\end{thebibliography}
\end{document}